\documentclass{aa}
\usepackage{graphicx}
\usepackage{natbib}
\usepackage{color}
\usepackage{fancyvrb} 
\usepackage{amssymb}
\usepackage{bm,url,times}
\usepackage{enumerate}
\graphicspath{{./fig/}{./ps/}}

\newcommand{\Fig}[1]{Fig.~\ref{#1}}
\newcommand{\Tab}[1]{Table~\ref{#1}}
\newcommand{\dd}{{\rm d} {}}
\newcommand{\EQ}{\begin{equation}}
\newcommand{\EN}{\end{equation}}
\newcommand{\EQA}{\begin{eqnarray}}
\newcommand{\ENA}{\end{eqnarray}}

\newcommand{\Sec}[1]{Sect.~\ref{#1}}

\newcommand{\setA}{MDI360} 
\newcommand{\setB}{HMI360}
\newcommand{\setC}{MDI72}
\newcommand{\setD}{HMI72}
\def\Rs{R_{\odot}}

\title{Solar-cycle variation of the rotational shear near the
  solar surface}  
                                   
\author{A. Barekat\inst{1} \and J. Schou\inst{1} \and L. Gizon\inst{1,2}}

\date{\today}

\institute{
Max-Planck-Institut f\"ur Sonnensystemforschung, Justus-von-Liebig-Weg
3, 37077 G\"ottingen, Germany
\and
Institut f\"ur Astrophysik, Georg-August-Universit\"at G\"ottingen,
37077 G\"ottingen, Germany
}
\begin{document}

\abstract
{
Helioseismology has revealed that the angular velocity of the Sun
increases with depth in the outermost 35 Mm of the Sun. Recently, we
have shown that  
the logarithmic radial gradient ($\dd\ln\Omega/\dd\ln r $) in the
upper 10~Mm is close to $-1$ from
the equator to $60^\circ$ latitude. 
}
{
We aim to measure the temporal variation of the
rotational shear over solar cycle 23 and the rising phase of cycle 24
(1996-2015). 
}
{
We used f mode frequency splitting data spanning 1996 to 2011 from
the Michelson Doppler Imager (MDI) and 2010 to 2015 from the
Helioseismic Magnetic Imager (HMI). In a first for such studies, the f
mode frequency splitting data were obtained from 360-day time series.
We used the same method as in our previous work for measuring
$\dd\ln\Omega/\dd\ln r $ from the equator to $80^\circ$ latitude in
the outer 13~Mm of the Sun. Then, we calculated the variation of
the gradient at annual cadence relative to the average over 1996 to 2015.  
}
{
We found the rotational shear at low latitudes ($0^\circ$ to
$30^\circ$) to vary in-phase with the solar activity, varying by  
$\sim \pm 10$\% over the period 1996 to 2015. At high latitudes
($60^\circ$ to $80^\circ$), we found rotational shear to vary in
anti-phase with the solar activity. 
By comparing the radial gradient obtained from the
splittings of the 360-day and the corresponding 72-day time series of
HMI and MDI data, 
we suggest that the splittings obtained from the 72-day HMI time
series suffer from systematic errors.
}
{
We provide a quantitative measurement of the temporal variation of the
outer part of the near surface shear layer which may provide useful
constraints on dynamo models and differential rotation theory.
}

\maketitle
\section{Introduction}
\label{introduction}

One of the major challenges in solar physics is to understand the
physics behind the 11-year solar cycle. In many
dynamo models, which attempt to explain the solar cycle,
 the differential rotation of the Sun plays an important role (see
the reviews by \cite{AxelK05} and \cite{Charb10}). 

In an $\alpha\Omega$ dynamo, rotational shear is
responsible for the $\Omega$-effect which 
generates toroidal magnetic field from a poloidal magnetic field.
The time variation of the shear has a
direct influence on the magnetic field generation in the Sun
as it may provide non-linear feedback on
the dynamo mechanism \citep{KAR99}. 
Additionally, the radial shear in the near-surface shear layer is
a potential explanation for the equatorward migration of the
activity belt during the 
solar cycle \citep{Axel05}. Hence, providing quantitative
information about the radial gradient of the rotation close to the
surface of the Sun is indispensable.  
Measurements of the radial shear can also deliver
constraints on differential rotation models \citep[e.g.,][]{KiRu05}.
\cite{Ki16} recently related the near-surface shear to the
subsurface magnetic field. Therefore, the time
variation of the shear with the solar cycle 
may also help estimate the strength of
the magnetic field below the surface at different phases of the cycle.

The radial shear can be measured by
several helioseismic techniques; see \cite{Thompson96}, 
\cite{JS98}, and the 
latest reviews of global and local 
helioseismology by \cite{Howe09} and \cite{GBS10}, respectively.
\cite{CT02} showed that the logarithmic
radial gradient in the outer 16~Mm of the Sun is
close to $-1$ up to $30^\circ $ latitude and becomes positive
above $55^\circ$ latitude. However,
\cite{BSG}, hereafter BSG, found no indication
of a change of sign at this latitude.   

\cite{AnBa08} studied the time variation of the radial and
latitudinal shear during solar cycle 23. They used 12 years
 (1996-2007) of p mode and f mode frequency splitting data from the Michelson  
Doppler Imager \citep[MDI;][]{Scher95} on board the Solar and
Heliospheric Observatory (SOHO). They also used 13 years (1995-2007) 
p mode frequency splitting data from the Global Oscillation Network
Group (GONG). They 
applied a two-dimensional regularized least square method \citep{AnBa98}
for inferring 
the rotation rate. Then, they studied the time variation of both
radial and latitudinal shears at several depths and
  latitudes. They found that the   
variation of the radial shear is about $20\%$ of its
average value at low latitudes at 14 Mm and below. 

\begin{table*}[!t]\caption{Summary of 15 years of the MDI and five years (16-20)
   of the HMI data.} 
\vspace{12pt}\centerline{\begin{tabular}{c  c c c c c c c c c c c c c
      c c}                             
\hline
data set & starting date & $n$&$l$& $n_1$&$l_1$&
$n_2$&$l_2$ &$n_3$&$l_3$ & $n_4$&$l_4$&
$n_5$&$l_5$ & $n_{\rm c}$&$l_c$\\ 
\hline
1 & 1996.05.01 & 187&98 & 128&134 & 123&124 & 122&142  &143&126  & 132&134 & 85&180 \\
2 & 1997.04.26 & 182&104& 120&133 & 129&135 &  126&136 & 128&131 & 129&121 & 83&175 \\
3 & 1998.04.21 & 168&114& 129&125 & 113&134 & 128&130 &- &- &- &- &- &-\\
4 & 1999.04.16 & 160&111& 132&127 & 134&133 & 127&135 & 127&135 & 128&127 & 86&164\\
5 & 2000.04.10 & 181&86 & 133&136 & 134&123 & 129&123 & 132&117 & 132&121 & 86&177\\
6 & 2001.04.05 & 177&108& 127&126 & 124&147 & 133&113 & 132&127 & 122&140 & 76&191\\
7 & 2002.03.31 & 179&87 & 130&144 & 121&121 & 131&137 & 129&144 & 130&124 & 83&181\\
8 & 2003.03.26 & 179&103& 125&140 & 119&144 & 134&133 & 116&124 & 131&111 & 79&176\\
9  & 2004.03.20 & 178&96& 119&138 & 124&138 & 126&116 & 119&146 & 131&129& 70&180 \\
10 & 2005.03.15 & 178&106& 131&134 & 129&135 & 128&127 & 128&118 & 124&132 & 75&180\\
11 & 2006.03.10 & 173&95& 132 &137 & 126&134 & 133&127 & 123&119 & 131&130 & 77&166\\
12 & 2007.03.05 & 177&106& 137&130 & 125&121 & 130&132 & 124&139 & 126&123 & 80&175\\
13 & 2008.02.28 & 184&96& 132&125 & 135&118 & 140&120 & 137&141 & 123&134 & 79&174\\
14 & 2009.02.22 & 183&91& 140&135 & 131&122 & 127&141 & 126&136 & 125&118 & 70&172\\
15 & 2010.02.17 & 163&128& 135&134 & 138&136 & 126&139 & 128&133 & 113&138 & 84&164\\ 
16 & 2010.04.30 & 149&131& 119&132 & 115&139 & 125&123 & 120&157 & 109&144 & 74&173\\
17 & 2011.04.25 & 152&118& 119&127 & 116&135 & 122&140 & 120&136 & 127&133 & 79&184\\
18& 2012.04.19 & 152&110& 115&144 & 116&131 & 117&137 & 114&130 & 108&161 & 70&196\\
19 & 2013.04.14 & 166&100& 125&128 & 118&147 & 124&140 & 124&141 & 118&134 & 77&171\\
20 & 2014.04.09 & 173&103& 119&149 & 122&149 & 114&138 & 126&125 &
119&142 & 75&180\\
\hline
\label{MDI-HMI}\end{tabular}}
\tablefoot{The number of splitting frequencies available
  in each data set is given by $n$ and $n_1$ to $n_5$ are the
  constituent 72-day time series. The minimum angular
  degree recovered is given by $l$ in 
  each set. $n_c$ and $l_c$ represent similar quantities for the
  common modes between each 360-day and the corresponding 
  72-day data sets. For details see \Sec{svs.t}.} 
\end{table*}

In this work, we investigate the solar cycle variation of the 
radial gradient of the rotation in the
outer 13~Mm of the Sun using f modes. 
We use 19 consecutive years of frequency splitting 
data corresponding to the entire solar cycle 23 (1996-2010) and the
rising phase of cycle 24 (2010-2015).
These data are obtained from 360-day time series from the Medium-$l$
program of MDI
and from the Helioseismic and Magnetic Imager 
\citep[HMI;][]{Schou12} on board the Solar Dynamics Observatory. These
data are different from what we used in BSG
in which the splittings were obtained from 72-day time
series. Therefore, we compare the gradient obtained from these two
different data sets in \Sec{svs.t} before we investigate the time
variation of the gradient in \Sec{tvs}.

\section{Observational data}
\label{OBS}

We consider only f modes. We denote mode frequency by $\nu_{lm}$
where $l$ and $m$ are the spherical 
harmonic degree and for azimuthal order, respectively.
We use 18 odd $a$-coefficients for each $l$ \citep{JCT94} obtained from 
MDI and HMI data, which are defined by 
\EQ
\nu_{lm}= \nu_{l} + \sum_{j=1}^{36}a_{l,j}\mathcal{P}_j^{(l)}(m),
\EN
where $\nu_{l}$ is the mean multiplet frequency, and
$\mathcal{P}_j^{(l)}$ are orthogonal polynomials of degree $j$. 
We use two sets of data of 
each instrument; the $a$-coefficients which are obtained from 72-day
and 360-day time series, resulting in four data sets:
\begin{itemize}
\item \setA: 15 sets obtained from  360-day  MDI (1996-2011) 
\item \setB: 5 sets obtained from   360-day HMI (2010-2015)
\item \setC: 74 sets obtained from  72-day MDI (1996-2011)
\item \setD: 25 sets obtained from  72-day  HMI (2010-2015).
\end{itemize} 
We summarize the number of modes found
in each data set in \Tab{MDI-HMI}. 
The differences between the splittings obtained from 360-day and
72-day time 
series of MDI data were investigated in great detail by \cite{LS15},
who also provide further details on the analysis.

 The \setC~and \setD~are used only for the comparison
between the results obtained from these data sets and the
corresponding results obtained from data sets \setA~and \setB. 
We note here that each 360-day time series is the 
combination of the five corresponding 72-day ones except for the third data set in
\Tab{MDI-HMI}, which was 
made from three non-consecutive 72 day time series \citep{LS15}
because of problems with the SOHO spacecraft.

\section{Method}

Our method for measuring the radial gradient is identical to the
one used by BSG. We explain our method here
succinctly and refer the reader to BSG for detailed explanation. 
We model the rotation rate as changing linearly with 
depth
\EQ
\Omega(r,u)=\Omega_0(u)+(1-r)\Omega_1(u),
\label{omega}
\EN
where $r$ is the distance to the center of the Sun  normalized by its
photospheric radius ($\Rs$), $u$ is the cosine of co-latitude and,
$\Omega_0(u)$ 
and $ \Omega_1(u)$ are the rotation rate at the surface and the slope,
respectively. Then, we perform a forward problem using the relation
between the $a$-coefficients and $\Omega$ which is given by
\EQ
2\pi a_{l,2s+1} = \int_0^1  \dd r \int_{-1}^1 \dd u \; K_{ls}(r,u)\Omega(r,u),
\label{a-coff}
\EN
 where $K_{ls}$ are kernels. We obtain
\EQ
\widetilde{\Omega}_{ls} \equiv {2\pi a_{l,2s+1}\over \beta_{ls} } =
\langle\Omega_0\rangle_s + (1-\overline{r}_{ls}) \langle
\Omega_1 \rangle_s,
\label{fittt}
\EN
where the $\beta_{ls}$ are the total integrals of the radial component
of the kernels (see Eq.(4) in BSG) and 
$\overline{r}_{ls}$ is the central of gravity of the radial
kernels. The $\langle \, 
\rangle$ denotes latitudinal averages. 
Next, we perform an error-weighted least square fit of
$\widetilde{\Omega}_{ls}/2\pi$ versus $(1-\overline{r}_{ls})$ to
determine $\langle \Omega_0 \rangle_s$ and $\langle \Omega_1
\rangle_s$ for each data set. 

In the last step of our analysis, we apply the inversion method
used by \cite{JS99} to $\langle\Omega_0 \rangle_s$ and $\langle \Omega_1 \rangle_s$ to infer the
rotation rate at each target latitude $u_0$ and from this obtain
$\dd\ln\Omega/\dd\ln r$.

\section{Results}

\begin{figure}[t!]
\begin{center}
\includegraphics[width=\columnwidth]{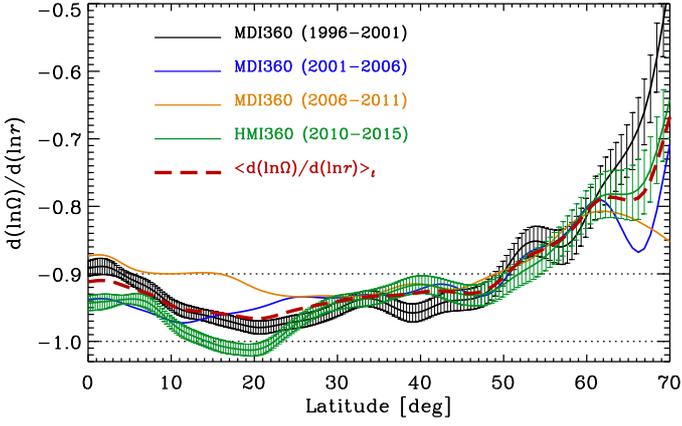}
\end{center}\caption[]{Time average of the
  logarithmic radial gradient versus target latitude. Black, blue,
  and orange lines represent each consecutive five year time average of
  $\dd\ln\Omega/\dd\ln r$ obtained from \setA. The green line shows
  the same quantity obtained from \setB. The red dashed line shows
  the 19 year (1996-2015) time average of $\dd\ln\Omega/\dd\ln r$.       
   The error bars are $1\sigma$. The errors on the orange and blue
   lines are similar to the black one. The errors on the red dashed line are
    similar to the thickness of the line.
}\label{tav}\end{figure}

Figure 1 shows the radial gradient obtained from the \setA~and the
\setB~data sets. 
Also shown in \Fig{tav} and summarized in \Tab{tab-hm}
is the value of the 19 year (1996-2015) time average of
$\dd\ln\Omega/\dd\ln r$. 
Going from the equator, this average fluctuates between $-0.97$ and
$-0.9$ up to $50^\circ$ latitude, above which it steadily increases
with latitude.
We included data sets 15 and 16 in the
average even though they have 288 days of overlap. 

Figure 1 also shows consecutive five year time averages of  
$\dd\ln\Omega/\dd\ln r$ which roughly represent different phases of
two solar cycles. There is evidence of the solar cycle
variation of $\dd\ln\Omega/\dd\ln r$ at low and high latitudes. These results
lead us to investigate the temporal variation of
$\dd\ln\Omega/\dd\ln r$ with annual cadence. We show the results in
\Sec{tvs}.   

We note that the time averaged value obtained from 
the \setB~ data set does not show the same trend as the one measured
in BSG above 
$60^\circ$ latitude using the first 20 sets of \setD~(see \Fig{fit} of
BSG).  
We explore the difference between our results and BSG of each
instrument in detail in the next section. 

\begin{table}[!t]\caption{Selected values of 19 year (1996-2015) time
    averaged values of the logarithmic radial gradient from \Fig{tav}.  
}\vspace{12pt}\centerline{\begin{tabular}{cc}                            
\hline \hline
Latitude  & $<\dd\ln\Omega/\dd\ln r>_t$ \\
\hline
$0^\circ$   & $-0.912\pm 0.004$ \\ 
$10^\circ$  & $-0.947\pm 0.003$ \\ 
$20^\circ$  & $-0.966\pm 0.004$ \\
$30^\circ$  & $-0.941\pm 0.004$ \\
$40^\circ$  & $-0.927\pm 0.005$ \\
$50^\circ$  & $-0.906\pm 0.007$ \\
$60^\circ$  & $-0.809\pm 0.011$ \\
\hline              
\label{tab-hm}\end{tabular}}
\end{table}

\subsection{Results obtained from 72-day vs. 360-day data}
\label{svs.t}

In this section, we compare the radial gradient derived from splittings
obtained from 360-day time series and 72-day time series from both MDI
and HMI. First, we show the results of MDI data and then
HMI.

\begin{figure}[t!]
\begin{center}
\includegraphics[width=\columnwidth]{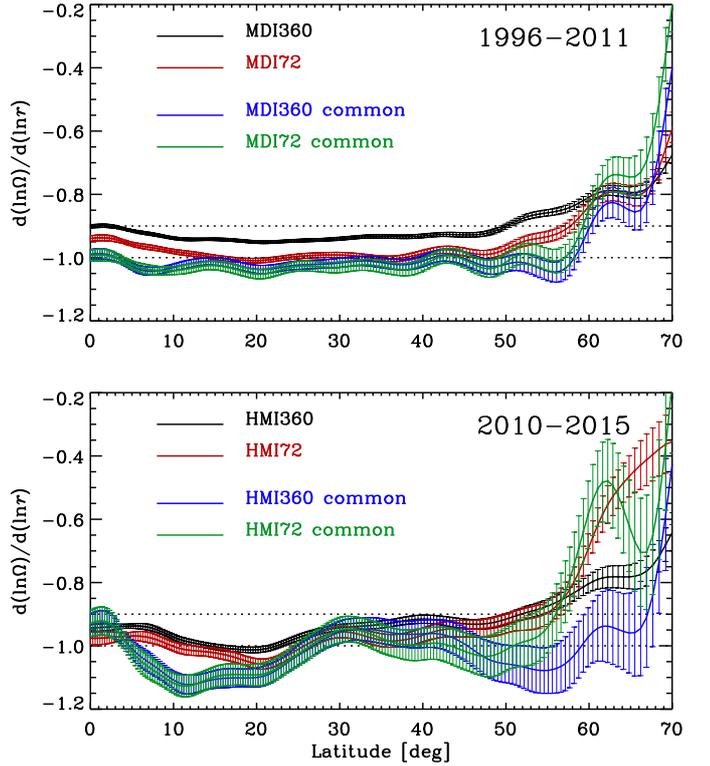}
\end{center}\caption[]{Comparison of time averages of
  $\dd\ln\Omega/\dd\ln r$ versus target latitude using 15 years of MDI data
  (upper panel) and five years of HMI data (lower panel). In both panels, black
  and blue lines show results obtained from splittings from 360-day
  time series of all and common modes (see, \Sec{svs.t}),
  respectively. The red and green lines show the results obtained from 
  72-day time series of all and common modes, respectively. The dotted
  lines mark the constant values of $-0.9$ and $-1$ at all
  latitudes. The error bars are $1\sigma$.  
}\label{MMHH}\end{figure}

\begin{figure}[t!]
\begin{center}
\includegraphics[width=\columnwidth]{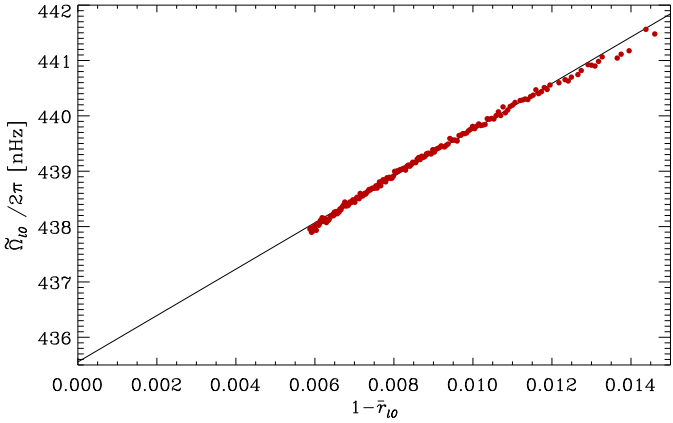}
\end{center}\caption[]{$\widetilde{\Omega}_{l0}/2\pi$ verses
  $(1-\overline{r}_{l0})$ obtained from the data set starting 10 April 
  2000 of \setA. The black line is the error weighted linear least square fit.
  We avoid plotting the error bars as they are in similar size of the symbols.
}\label{fit}\end{figure}

The first panel of \Fig{MMHH} shows the 15 year (1996-2011) time
average of the radial gradient obtained from data sets \setA~and
\setC. 
The result from data set \setC~is identical to the MDI result found by
BSG.
For the MDI data, the absolute value of $\dd\ln\Omega/\dd\ln r$ is about 5\% 
smaller than the values found by BSG. This difference can be explained
by the fact that using \setA~and \setB~data sets enables us to probe
roughly 3 Mm 
deeper than using data sets \setC~and \setD. As a consequence, 
 $\widetilde{\Omega}_{l0}/2\pi$ is not linear in $r$ any more,
as shown in \Fig{fit}, which in turn means that the fitted values
depend on the modes included.

The maximum value of $l=300$ is the same for all data sets, but the
minimum value of $l$ is different; see \Tab{MDI-HMI}. Therefore,
we compare the results obtained from
each set in \setA~with those from the corresponding
five sets of \setC, using only the common modes. 
The result is shown in the first panel of
\Fig{MMHH}. For this comparison we excluded the last data set of
\setC~because it is after the last set in \setA. 
Considering only common modes causes us
to exclude more than 
half of the modes from each data set in \setA~(see last column of
\Tab{MDI-HMI}). The difference between the results obtained from \setA~and
\setC~are reduced substantially  
and they are now in agreement to better than $1\sigma$ up to $50^\circ$
latitude. This difference increases gradually toward higher
latitudes which shows that the results above $50^\circ$ latitude are not
reliable. We note that one would expect the results to be consistent to
better than $1\sigma$, as they are obtained from the same underlying data.
Thus there is clear evidence that the splitting data suffer from
systematic errors. 

We applied the same comparison to sets \setB~and \setD. The five year time
averages from using both all and only the common modes are shown in
the bottom panel of \Fig{MMHH}.   
There is a significant discrepancy between the two
results obtained from sets \setB~and \setD~above $60^\circ$ latitude
which does not disappear even when comparing the results obtained from
the common modes.  
This shows that the HMI data are even more affected by systematic
errors than the MDI data. 

For HMI data, we carry out further analysis by comparing the results
derived from common modes of each year. Except for
the first and last year the difference between the results
persists. The perfect agreement of the results in the last year
encourage us to compare common modes between these two data sets.
This comparison shows that the difference between $a_3$ and $a_5$ of
those data sets are significant. In average, the values of $a_3$ of
\setB~ is larger and $a_5$ is smaller than the corresponding \setD~
ones by about $3\sigma$. There are also clear systematic 
errors in those coefficients with larger discrepancies in the earlier 
than in the later years. 

Unfortunately, these comparisons do not tell us what causes the
systematic errors or how to correct them.
Understanding this will require a more detailed analysis (Larson \&
Schou in prep.). However, our results suggest that \setD~suffer from
systematic errors 
as the results obtained using \setB~are not significantly
different from the results of data sets \setA~ and
\setC. Moreover, we expect that the splittings obtained
from longer time series have better quality as the peaks are better
resolved. 
\citep{LS15}. 

\begin{figure}[t!]
\begin{center}
\includegraphics[width=\columnwidth]{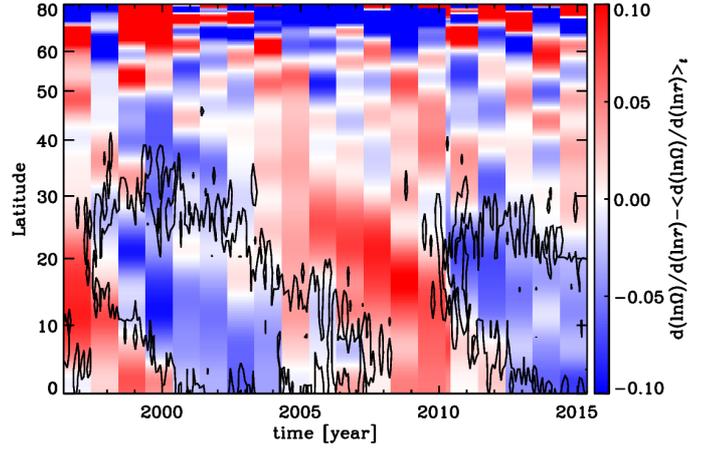}
\end{center}\caption[]{
Time variation of $\dd\ln\Omega/\dd\ln r$ relative to its 19 year time
average. The thin stripe in the plot shows the result obtained from
data set 15 overploted on data set 16 as there is 288 days overlap
between these two data sets.
The contours show the two hemisphere
averaged butterfly diagram of the sunspot area of 5 per millionths of a
hemisphere (courtesy of D. Hathaway; see
\url{http://solarscience.msfc.nasa.gov/greenwch.shtml}). 
}\label{tv}\end{figure} 

\subsection{Solar cycle variation of the radial gradient}
\label{tvs}

We measure the variation of $\dd\ln\Omega/\dd\ln r$ relative to its
time averaged value from 1996 to 2015 using data sets \setA~and
\setB. We show the results in \Fig{tv} together with the butterfly
diagram. These measurements reveal two cyclic patterns; 
 one at low latitudes from the equator to about $40^\circ$
latitude and one above $60^\circ$ latitude. There is no clear signal
between about $40^\circ$ and $60^\circ$ latitude.

Below $40^\circ$, there exist bands where the rotation
  gradient is
  about 10\% larger and smaller than the average. As illustrated by
  the butterfly diagram in Figure 4, the band with steeper than average
  gradient (blue in \Fig{tv}) follows the activity belt quite
  closely. These bands are 
  also similar to the torsional oscillation signal (see,
  e.g., \cite{Howe06,AnBa08})

The temporal variation of $\dd\ln\Omega/\dd\ln r$ at high latitudes is
more than 10\% of its average value and has the opposite behavior to that
at low latitudes.
However, as we pointed out eirlier the measured values of the
gradient above $50^\circ$ latitude are not reliable, so any results
here have to be interpreted with caution.

\begin{figure}[t!]
\begin{center}
\includegraphics[width=\columnwidth]{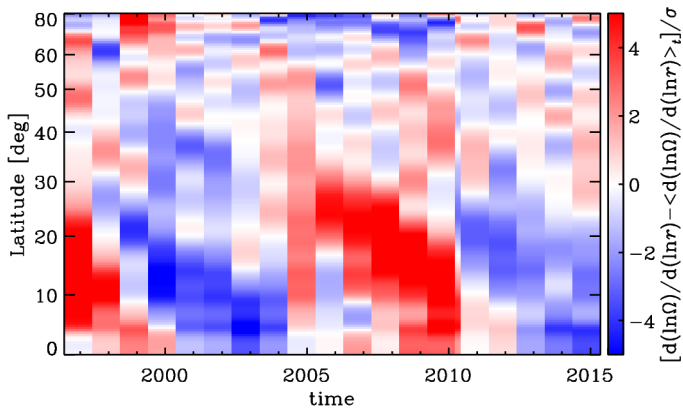}
\end{center}\caption[]{
Statistical significance in change of $\dd\ln\Omega/\dd\ln r$ relative to its
average values at different latitudes and time. $\sigma$ is the error
on $\dd\ln\Omega/\dd\ln r$ of each year.     
}\label{sig}\end{figure}

\begin{figure}[t!]
\begin{center}
\includegraphics[width=\columnwidth]{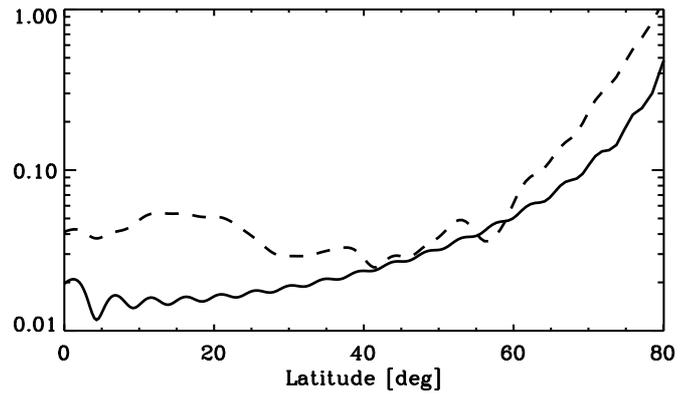}
\end{center}\caption[]{
Comparison between the standard deviation of the time variation of the
radial shear relative to its time averaged value (dashed line) and
the time averaged error of the shear (solid line) obtained
from data sets \setA~and \setB.
}\label{ersd}\end{figure}

The statistical significance of these signals is shown in \Fig{sig}
and the standard deviation in time 
and the time averaged errors in \Fig{ersd}. The measured
signals are statistically significant at low and high latitudes
as they are at the 3 to 8 $\sigma$ level, while they are indeed
not significant between $40^\circ$ and $60^\circ$ latitude.

  We note here that the results obtained from MDI360 and HMI360
  are only different by about 1\% when using modes with $l
\geqslant 120$, 
  corresponding roughly to the range used by the 72-day analysis and over
  which the rotation rate changes linearly with depth.  

It is well known that the phase and amplitude of the solar
  cycle variations of the rotation rate vary with depth and latitude
  \citep{VSCT02,AnBa03,Howe05,AnBa08}, but the
  temporal variation 
  of the gradient has not been previously reported over the same depth
  range as used in this work.

  \cite{AnBa08} found a similar pattern with similar amplitude of the
  temporal variation of 
  the radial gradient at $0.98\Rs$ as ours. They used the first eight odd $a$-coefficients
  obtained from MDI p and f modes 
   and GONG p modes spanning 1995 to 2007.
  Despite the similarity in pattern and amplitude, the
  sign of the change in the gradient of their results is opposite to
  ours. They saw that sunspots  
  occurred where the absolute value of the gradient is smaller than
  the average value which is the opposite of what we see.
  This difference might come from the
  fact that we are measuring the temporal variation around $0.99\Rs$ while
  they measured it at $0.98\Rs$. 
  We also note that \cite{AnBa08} used the
  earlier version of the MDI data (see \cite{LS15} and BSG)
  which might explain the discrepancy that
  \cite{AnBa08} saw between GONG and MDI data
  at $0.98\Rs$ and shallower layers. 
  
\section{Conclusion}

We make measurements of the radial gradient over 19 years (1996-2015)
corresponding to solar cycle 23 and the rising phase of cycle 24 in the
outer 13~Mm of the Sun. We use recently available f
mode frequency splittings data obtained from 360-day time series of
MDI spanning 1996 to 2011 and HMI spanning 2010 to 2015. The
values of the radial gradient derived from \setA~and \setB~fluctuate between
$-0.97$ and $-0.9$ up to $50^\circ$ latitude. These values are a
few percent larger than 
measured values by BSG which are obtained from \setC~and \setD. It
turns out that this difference comes from 
the fact that the angular velocity does not change linearly with depth
to deeper than about 10~Mm below the surface.  

We also compare the radial gradient
obtained from common modes of two different data sets of each
instrument. These comparisons reveal that the measured values of 
$\dd\ln\Omega/\dd\ln r$ above $50^\circ$ latitude are not
reliable. Another important finding is that there are considerable
systematic errors in HMI data that needs further investigation.

By measuring the variation of rotational shear relative to its
19 year time averaged value we find two cyclic patterns at low
($0^\circ$ to $30^\circ$) and at high ($60^\circ$ to $80^\circ$)
latitudes with similar period of the solar cycle. 
Both patterns show bands of larger and smaller than average shear
moving toward the equator and poles at low and high latitudes,
respectively. The relative change in the shear is about 10\% at low
latitudes and 20\% at high latitudes. Although the values of
$\dd\ln\Omega/\dd\ln r$  
above $50^\circ$ are not reliable, the temporal variation of
$\dd\ln\Omega/\dd\ln r$ is significant above $60^\circ$ latitudes.
This finding may have important   
implications for dynamo models as this variation is considerable compared
to the torsional oscillation \citep{AnBa08}.

The cyclic behavior
of the shear at low latitudes agrees with the recent 
theoretical work by \cite{Ki16} who showed that the strength of the
shear increases because of the presence of the strong magnetic
field. Therefore accurate measurements of the shear 
might be a way of determining of the sub-surface magnetic field.

\begin{acknowledgements}
We thank T.~P. Larson for discussions regarding details of the HMI
data, and A. Birch for various discussions and his useful comments
about the paper.
L. Gizon acknowledges support from the Center for Space Science at the
NYU Abu Dhabi Institute.
SOHO is a project of international cooperation between ESA and NASA.
The HMI data are courtesy of NASA/SDO and the HMI science team.
\end{acknowledgements}

\bibliographystyle{aa}
\bibliography{all}

\newcommand{\noop}[1]{}
\begin{thebibliography}{23}
\expandafter\ifx\csname natexlab\endcsname\relax\def\natexlab#1{#1}\fi

\bibitem[{{Antia} {et~al.}(1998){Antia}, {Basu}, \& {Chitre}}]{AnBa98}
{Antia}, H.~M., {Basu}, S., \& {Chitre}, S.~M. 1998, \mnras, 298, 543

\bibitem[{{Antia} {et~al.}(2008){Antia}, {Basu}, \& {Chitre}}]{AnBa08}
{Antia}, H.~M., {Basu}, S., \& {Chitre}, S.~M. 2008, \apj, 681, 680

\bibitem[{{Barekat} {et~al.}(2014){Barekat}, {Schou}, \& {Gizon}}]{BSG}
{Barekat}, A., {Schou}, J., \& {Gizon}, L. 2014, \aap, 570, L12

\bibitem[{{Basu} \& {Antia}(2003)}]{AnBa03}
{Basu}, S. \& {Antia}, H.~M. 2003, \apj, 585, 553

\bibitem[{{Brandenburg}(2005)}]{Axel05}
{Brandenburg}, A. 2005, \apj, 625, 539

\bibitem[{{Brandenburg} \& {Subramanian}(2005)}]{AxelK05}
{Brandenburg}, A. \& {Subramanian}, K. 2005, \physrep, 417, 1

\bibitem[{{Charbonneau}(2010)}]{Charb10}
{Charbonneau}, P. 2010, Living Reviews in Solar Physics, 7, 3

\bibitem[{{Corbard} \& {Thompson}(2002)}]{CT02}
{Corbard}, T. \& {Thompson}, M.~J. 2002, \solphys, 205, 211

\bibitem[{{Gizon} {et~al.}(2010){Gizon}, {Birch}, \& {Spruit}}]{GBS10}
{Gizon}, L., {Birch}, A.~C., \& {Spruit}, H.~C. 2010, \araa, 48, 289

\bibitem[{{Howe}(2009)}]{Howe09}
{Howe}, R. 2009, Living Reviews in Solar Physics, 6

\bibitem[{{Howe} {et~al.}(2005){Howe}, {Christensen-Dalsgaard}, {Hill}, {Komm},
  {Schou}, \& {Thompson}}]{Howe05}
{Howe}, R., {Christensen-Dalsgaard}, J., {Hill}, F., {et~al.} 2005, \apj, 634,
  1405

\bibitem[{{Howe} {et~al.}(2006){Howe}, {Komm}, {Hill}, {Ulrich}, {Haber},
  {Hindman}, {Schou}, \& {Thompson}}]{Howe06}
{Howe}, R., {Komm}, R., {Hill}, F., {et~al.} 2006, \solphys, 235, 1

\bibitem[{{Kitchatinov}(2016)}]{Ki16}
{Kitchatinov}, L.~L. 2016, Astronomy Letters, 42, 339

\bibitem[{{Kitchatinov} \& {R{\"u}diger}(2005)}]{KiRu05}
{Kitchatinov}, L.~L. \& {R{\"u}diger}, G. 2005, Astronomische Nachrichten, 326,
  379

\bibitem[{{K{\"u}ker} {et~al.}(1999){K{\"u}ker}, {Arlt}, \&
  {R{\"u}diger}}]{KAR99}
{K{\"u}ker}, M., {Arlt}, R., \& {R{\"u}diger}, G. 1999, \aap, 343, 977

\bibitem[{{Larson} \& {Schou}(2015)}]{LS15}
{Larson}, T.~P. \& {Schou}, J. 2015, \solphys, 290, 3221

\bibitem[{{Scherrer} {et~al.}(1995){Scherrer}, {Bogart}, {Bush}, {Hoeksema},
  {Kosovichev}, {Schou}, {Rosenberg}, {Springer}, {Tarbell}, {Title},
  {Wolfson}, {Zayer}, \& {MDI Engineering Team}}]{Scher95}
{Scherrer}, P.~H., {Bogart}, R.~S., {Bush}, R.~I., {et~al.} 1995, \solphys,
  162, 129

\bibitem[{{Schou}(1999)}]{JS99}
{Schou}, J. 1999, \apjl, 523, L181

\bibitem[{{Schou} {et~al.}(1998){Schou}, {Antia}, {Basu}, {Bogart}, {Bush},
  {Chitre}, {Christensen-Dalsgaard}, {Di Mauro}, {Dziembowski}, {Eff-Darwich},
  {Gough}, {Haber}, {Hoeksema}, {Howe}, {Korzennik}, {Kosovichev}, {Larsen},
  {Pijpers}, {Scherrer}, {Sekii}, {Tarbell}, {Title}, {Thompson}, \&
  {Toomre}}]{JS98}
{Schou}, J., {Antia}, H.~M., {Basu}, S., {et~al.} 1998, \apj, 505, 390

\bibitem[{{Schou} {et~al.}(1994){Schou}, {Christensen-Dalsgaard}, \&
  {Thompson}}]{JCT94}
{Schou}, J., {Christensen-Dalsgaard}, J., \& {Thompson}, M.~J. 1994, \apj, 433,
  389

\bibitem[{{Schou} {et~al.}(2012){Schou}, {Scherrer}, {Bush}, {Wachter},
  {Couvidat}, {Rabello-Soares}, {Bogart}, {Hoeksema}, {Liu}, {Duvall}, {Akin},
  {Allard}, {Miles}, {Rairden}, {Shine}, {Tarbell}, {Title}, {Wolfson},
  {Elmore}, {Norton}, \& {Tomczyk}}]{Schou12}
{Schou}, J., {Scherrer}, P.~H., {Bush}, R.~I., {et~al.} 2012, \solphys, 275,
  229

\bibitem[{{Thompson} {et~al.}(1996){Thompson}, {Toomre}, {Anderson}, {Antia},
  {Berthomieu}, {Burtonclay}, {Chitre}, {Christensen-Dalsgaard}, {Corbard}, {De
  Rosa}, {Genovese}, {Gough}, {Haber}, {Harvey}, {Hill}, {Howe}, {Korzennik},
  {Kosovichev}, {Leibacher}, {Pijpers}, {Provost}, {Rhodes}, {Schou}, {Sekii},
  {Stark}, \& {Wilson}}]{Thompson96}
{Thompson}, M.~J., {Toomre}, J., {Anderson}, E.~R., {et~al.} 1996, Science,
  272, 1300

\bibitem[{{Vorontsov} {et~al.}(2002){Vorontsov}, {Christensen-Dalsgaard},
  {Schou}, {Strakhov}, \& {Thompson}}]{VSCT02}
{Vorontsov}, S.~V., {Christensen-Dalsgaard}, J., {Schou}, J., {Strakhov},
  V.~N., \& {Thompson}, M.~J. 2002, Science, 296, 101

\end{thebibliography}

\end{document}